\documentclass[%
 reprint,
 amsmath,amssymb,
 aps,
]{revtex4-2}

\usepackage{graphicx}
\usepackage{dcolumn}
\usepackage{bm}

\begin{document}

\preprint{APS/123-QED}

\title{Spin-orbit-enhanced robustness of supercurrent in graphene/WS$_2$ Josephson junctions}

\author{T. Wakamura$^1$, N. J. Wu$^{1, 2}$, A. D. Chepelianskii$^1$, S. Gu\'{e}ron$^1$, M. Och$^3$, M. Ferrier$^1$, T. Taniguchi$^4$, K. Watanabe$^5$, C. Mattevi$^3$}
\author{H. Bouchiat$^1$}%
\email{helene.bouchiat@u-psud.fr}
\affiliation{$^1$
Universit\'{e} Paris-Saclay, CNRS, Laboratoire de Physique des Solides, 91405 Orsay Cedex, France}%

\affiliation{$^2$
Universit\'{e} Paris-Saclay, CNRS, Institut de Sciences Mol\'{e}culaires, 91405 Orsay Cedex, France}

\affiliation{$^3$Department of Materials, Imperial College London, Exhibition Road, London, SW7 2AZ, United Kingdom}%


\affiliation{$^4$International Center for Materials Nanoarchitectonics, National Institute for Materials Science, 1-1 Namiki, Tsukuba 305-0044, Japan}

\affiliation{$^5$Research Center for Functional Materials, National Institute for Materials Science, 1-1 Namiki, Tsukuba 305-0044, Japan}

\date{\today}

\begin{abstract}

We demonstrate the enhanced robustness of the supercurrent through graphene-based Josephson junctions in which strong spin-orbit interactions (SOIs) are induced. We compare the persistence of a supercurrent at high magnetic fields between Josephson junctions with graphene on hexagonal boron-nitride and graphene on WS$_2$, where strong SOIs are induced via the proximity effect. We find that in the shortest junctions both systems display signatures of induced superconductivity, characterized by a suppressed differential resistance at a low current, in magnetic fields up to 1 T. In longer junctions however, only graphene on WS$_2$ exhibits induced superconductivity features in such high magnetic fields, and they even persist up to 7 T. We argue that these robust superconducting signatures arise from quasiballistic edge states stabilized by the strong SOIs induced in graphene by WS$_2$.

\end{abstract}

\pacs{Valid PACS appear here}
\maketitle

Magnetic fields are known to be detrimental to ordinary s-wave superconductivity because of the pair-breaking effect of the Zeeman component, which can flip the spins in the spin-singlet Cooper pair \cite{Tinkham}. 
Magnetic fields also affect superconductivity via an orbital effect, a geometry-dependent dephasing of Cooper pairs by the vector potential. This orbital effect determines the field-dependent interference pattern of the critical current in spatially extended Josephson junctions \cite{Cuevas, Crouzy}. 

Spin-orbit interactions (SOIs) play a crucial role in mitigating these field-induced pair-breaking effects. Recently, Ising pairing in transition-metal dichalcogenides was found to confer the robustness of superconductivity owing to spin-momentum locking, by which the spin polarization of Cooper pairs is prevented \cite{Lu, Xi, Saito}. While the role of SOIs in stabilizing the spin component of the Cooper pair was emphasized in many previous studies, the effect of SOIs on orbital depairing is only beginning to be explored \cite{Zuo, Nijholt}.

In this Letter, we demonstrate that SOIs can enhance the superconducting proximity effect in high out-of-plane magnetic fields in graphene-on-WS$_2$-based Josephson junctions. These junctions consist of graphene encapsulated between hexagonal boron nitride (hBN) and WS$_2$, which induces the strong SOIs in graphene via the proximity effect \cite{ZWang1, ZWang2, Wakamura1, Wakamura2, Zihlmann, Cummings, David}. The magnetic field dependence of the critical current through graphene-based superconductor-normal-metal-superconductor junctions has been already extensively studied \cite{BenShalom, Calado, Borzenets}. It is characterized by a Fraunhofer-like pattern resulting from interference between the uniformly distributed Andreev pair trajectories, and decays rapidly at fields above a few flux quanta through the sample. However, specifically in very clean short ballistic junctions, residual supercurrent periodic oscillations at very high fields were observed and associated with the physics of the quantum Hall (QH) effect \cite{Amet}. Here, we investigate junctions in the opposite, diffusive limit. The junction lengths ($L$) range between 100 and 500 nm, from the short- to the long-junction regimes. Although the junctions are diffusive, surprisingly, we find clear signatures of induced superconductivity with manifestations of a supercurrent even in magnetic fields in the Tesla range for the graphene-on-WS$_2$ junctions. By contrast, this behavior is not observed for graphene-on-hBN junctions outside the short ballistic regime, i.e. for lengths greater than $L$ = 200 nm. We argue that this robust induced superconductivity arises from quasiballistic trajectories along the sample edges, stabilized by strong SOIs induced in graphene by WS$_2$.

\begin{figure}[tb]
\includegraphics[width=8.5cm,clip]{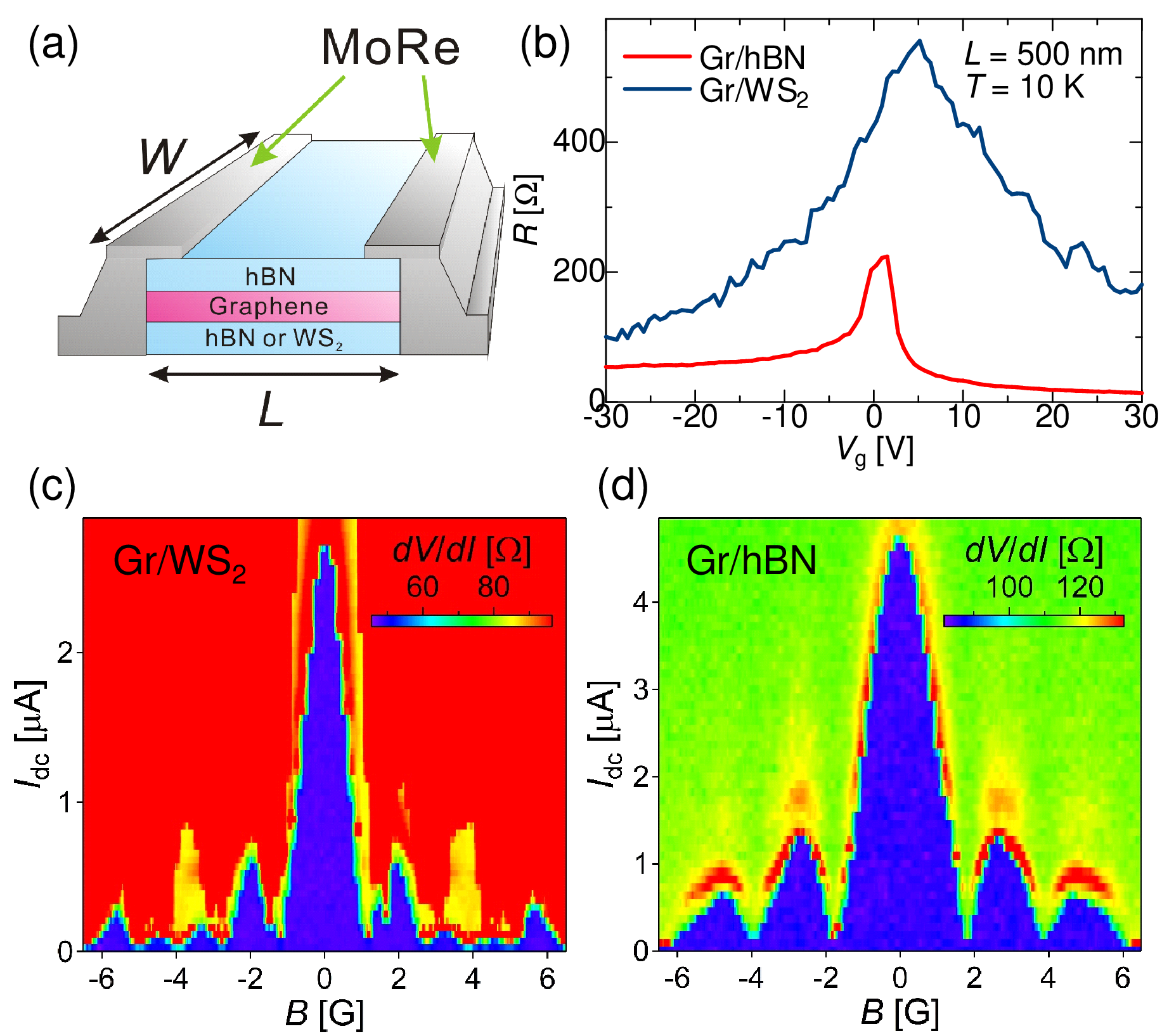}
\caption{(a) Schematic illustration of a graphene-based Josephson junction device of length $L$ and width $W$ employed in this study. (b) $V_g$ dependence of $R$ for the Gr/WS$_2$ and Gr/hBN junctions with MoRe in the normal state ($L$ = 500 nm). The contact resistance is subtracted in the data. (c) and (d) Color-coded $dV/dI$, plotted as a function of $I_{\rm dc}$ and $B$ for Gr/WS$_2$ [(c)] and Gr/hBN junctions [(d)] with $L$ = 500 nm measured at $V_g$ = 60 V.}
\label{Figure2}
\end{figure} 

\begin{figure*}[tb]
\begin{center}
\includegraphics[width=17.5cm,clip]{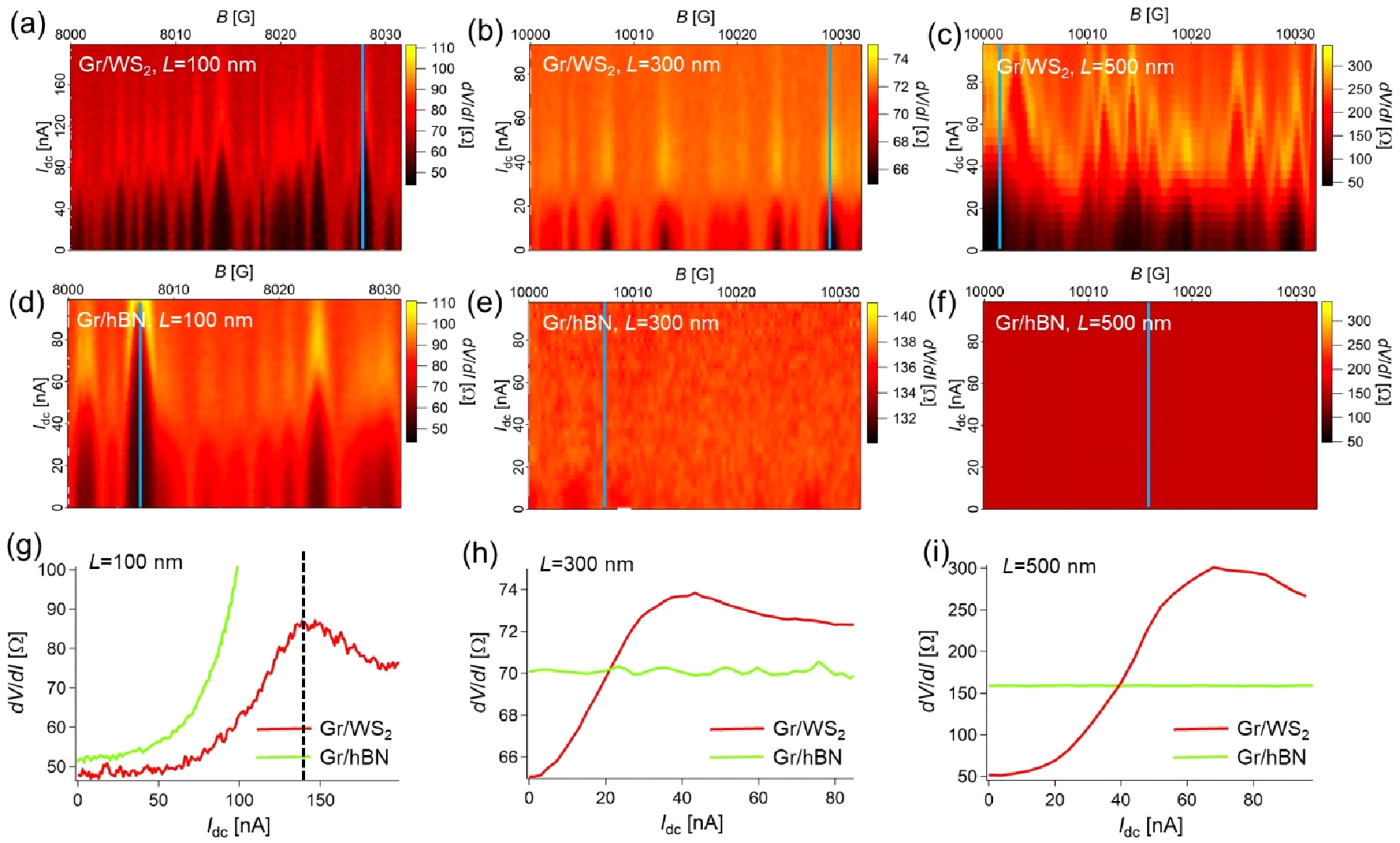}
\caption{Color-coded $dV/dI$ as a function $I_{\rm dc}$ and $B$ around $B$ = 10000 G at $V_g$ = 60 V for all samples. For $L$ = 100 nm ((a) and (d)), superconducting pockets are clearly visible, in the form of field regions of low $dV/dI$ at low $I_{\rm dc}$, for both the Gr/WS$_2$ and Gr/hBN junctions, around $B$ = 8000 G. For $L$ = 300 nm [(b) and (e)] and $L$ = 500 nm [(c) and (f)] , superconducting pockets are visible only for the Gr/WS$_2$ junction. (g)-(i) Cross-sectional image along the light blue line shown in (a)-(f) of $dV/dI$ as a function of $I_{\rm dc}$ for Gr/WS$_2$ and Gr/hBN junctions with different $L$. Red and light green curves are from Gr/WS$_2$ and Gr/hBN junctions, respectively. The suppressed $dV/dI$ at low $I_{\rm dc}$, signature of an induced superconducting proximity effect, is clearly visible for the Gr/WS$_2$ junctions of every length but only for the shortest Gr/hBN junction. In (g) the peak (or bump) of $dV/dI$ for Gr/hBN is located out of the range of $I_{\rm dc}$ in the measurement, and in (h)[(i)], the $dV/dI$ for Gr/hBN (Gr/WS$_2$) is vertically shifted to compare to that for Gr/WS$_2$ (Gr/hBN). The residual resistance around 50 $\Omega$ at $I_{\rm dc}$ = 0 for (a)-(i) arises from the measurement wires. The dashed line in (g) represents the value of $I_{\rm dc}$ which defines $I_c$.}
\label{fig2}
\end{center}
\end{figure*}

\begin{figure*}[tb]
\begin{center}
\includegraphics[width=17.5cm,clip]{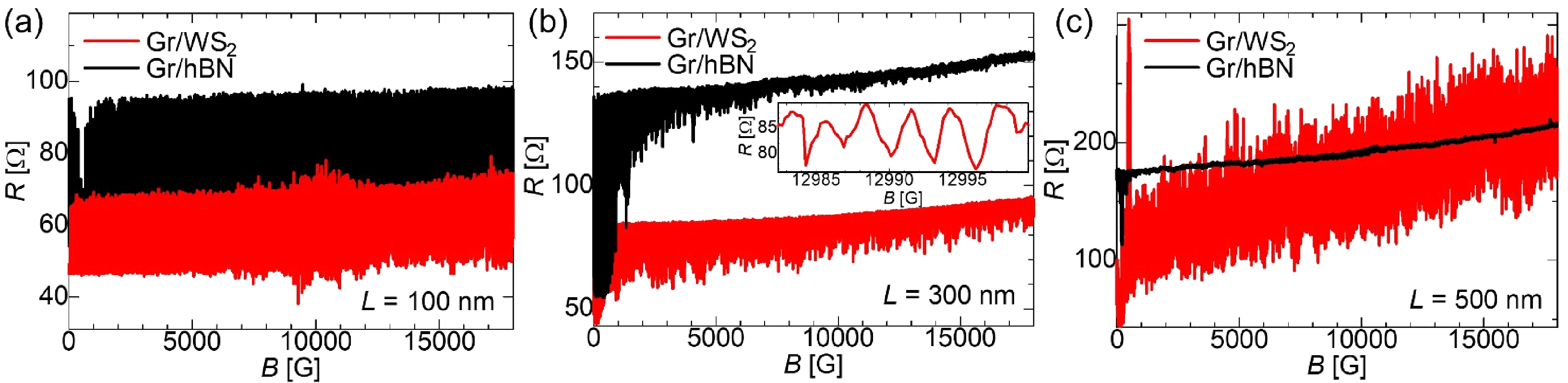}
\caption{Monitoring the superconducting proximity effect over a wide field range, via the zero bias differential resistance variations with $B$, for the three junction lengths and both Gr/WS$_2$ and Gr/hBN systems, at $V_g$= 60 V. 
(a) For $L$ = 100 nm, both Gr/WS$_2$ and Gr/hBN junctions display comparable oscillation amplitudes up to $B \sim$ 20000 G. (b) For $L$ = 300 nm, whereas the Gr/hBN junction displays larger amplitude oscillations near $B$ = 0, they are rapidly suppressed as $B$ increases. By contrast, the Gr/WS$_2$ junction displays a relatively large amplitude of resistance oscillations that persists even around 20000 G. (c) The difference between Gr/WS$_2$ and Gr/hBN is the most striking for $L$ = 500 nm junctions. The relative oscillation amplitude of the Gr/WS$_2$ junction's differential resistance is around 50 times greater than that of the Gr/hBN junction over the entire field range. The inset in (b) displays a magnified view of the oscillations for Gr/WS$_2$ around $B$ = 13000 G.}
\label{fig3}
\end{center}
\end{figure*}

We compare two types of samples: hBN/graphene/WS$_2$ (Gr/WS$_2$) and hBN/graphene/hBN (Gr/hBN) junctions. Graphene and hBN are mechanically exfoliated from graphite and hBN crystals, and monolayer WS$_2$ flakes are grown by chemical vapor deposition (CVD) \cite{Wakamura1, Reale}. hBN and graphene are picked up by the typical dry-transfer technique with polydimethylsiloxane and polypropylene carbonate and then deposited onto WS$_2$ or hBN \cite{LWang}. One-dimensional superconducting contacts are patterned by electron beam lithography, followed by reactive ion etching and sputtering 100 nm MoRe. MoRe is a type-I\hspace{-.1em}I superconductor with high critical field $H_{c2} \sim$ 80000 G (8 T) and critical temperature $T_{\rm c}$ $\sim$ 10 K \cite{Amet}. The junctions are defined by their length ($L$) and width ($W$), $W \sim$ 10 $\mu$m for all samples whereas $L$ varies between 100 and 500 nm [see Fig. 1(a)]. Measurements are performed in a dilution refrigerator, at 100 mK unless otherwise specified, using a conventional lock-in technique.

We previously demonstrated via weak antilocalization that graphene on TMDs acquires by proximity strong SOIs, thanks to the heavy elements such as molybdenum (Mo) or tungsten (W) that they contain \cite{Wakamura1, Wakamura2, Xiao}. While the intrinsic SOI in graphene is small (24 $\mu$eV) \cite{Gmitra}, it is enhanced by contact with a TMD flake, up to 1 meV $\sim$ 10 meV, depending on the type and thickness of the TMDs. Specifically, it was found that monolayer tungsten-based TMDs such as WS$_2$ or WSe$_2$ induce the strongest SOIs in graphene \cite{Wakamura1, Wakamura2}. Therefore, our Gr/WS$_2$ junctions include graphene with strong SOIs as a normal region. Since the resistivity of the TMDs is much larger than that of graphene, the electrical current can be considered to flow entirely through the graphene in the Gr/TMD bilayer. 

We first discuss the results in the normal state. Figure 1(b) displays a typical gate voltage ($V_g$) dependence of the resistance ($R$) for Gr/WS$_2$ and Gr/hBN junctions ($L$ = 500 nm) measured above the $T_{\rm c}$ of MoRe in zero field. The Dirac peak is sharper in the Gr/hBN junction than in the Gr/WS$_2$, indicating that the mobility of graphene on hBN is higher than that of graphene on WS$_2$. While it is reported that multilayered TMD flakes may constitute as flat and clean substrates for graphene as hBN \cite{Young}, we note that in our experiments we work with monolayer TMD grown by chemical vapor deposition, which may contain polymer residues left over from the transfer process to the sample substrates and which also is not as flat as the multilayer hBN used for the Gr/hBN junctions.
In fact, all the Gr/hBN junctions have higher mobility than the Gr/WS$_2$ junctions. Particularly, the Gr/hBN junction with $L$ = 100 nm displays oscillations of $R$ with $V_g$ in the hole-doped region, consistent with previous reports of Fabry-P\'{e}rot (FP) oscillations \cite{BenShalom, Calado, Borzenets}. This is also in agreement with our estimate of a mean free path $l_e \sim$ 100 nm from a diffusive sample whose $L$ is longer than $l_e$. Interestingly, FP oscillations are not observed in Gr/WS$_2$ junctions even for $L$ = 100 nm, consistent with the lower mobility of Gr/WS$_2$ junctions \cite{Supp}. 

After cooling the sample below $T_{\rm c}$ of the superconducting contacts, we measured the differential resistance ($dV/dI$) as a function of the dc current ($I_{\rm dc}$) and magnetic field ($B$) around zero field. A small ac current $I_{\rm ac}$ was added to $I_{\rm dc}$, and the corresponding ac voltage was detected by a lock-in amplifier, yielding $dV/dI$. Figure 1(c) and (d) displays the color-coded $dV/dI$ of Gr/WS$_2$ and Gr/hBN junctions [$L$ = 500 nm, same samples as Fig. 1(b)] as a function of $I_{\rm dc}$ and $B$ at low fields. Both samples exhibit clear regions with zero $dV/dI$ at a low dc current, corresponding to an induced supercurrent. Deviations from the typical Fraunhofer pattern are presumably due to current inhomonegeities through the junctions. We note that the critical current ($I_c$), defined by $I_{\rm dc}$ at which $dV/dI$ is maximum, is larger for the Gr/hBN junction than that for the Gr/WS$_2$ junction. This indicates that the induced superconductivity is stronger for the Gr/hBN junction at low fields, and is consistent with the higher mobility of the Gr/hBN junctions [see Fig. 1(b)].

The low field behavior displayed in Fig. 1(c), (d) is the expected Fraunhofer-like interference pattern for a supercurrent flowing uniformly throughout the entire width of the graphene sheet \cite{Tinkham, Cuevas, Crouzy}. Whereas the value of $I_c$ varies for different junctions, similar behaviors are observed for all junctions. The oscillation period corresponds to the junction area if the magnetic focusing effect is taken into account \cite{Supp, Sumionen, Gu}.

To investigate the induced superconductivity at high fields, we next increased $B$ around 10000 G and similarly measured $dV/dI$ as a function of $I_{\rm dc}$ and $B$. Figure 2 compares Gr/WS$_2$ and Gr/hBN junctions with different $L$ ($L$ = 100, 300 and 500 nm). Considering the relation between the Thouless energy ($E_T = \hbar D/L^2$ with the diffusion constant $D$ in the diffusive regime and $E_T = \hbar v_F/L$ in the ballistic regime) and the superconducting gap $\Delta_0$ (= 1 meV) of MoRe, both $L$= 100 nm junctions are in the short-junction limit ($E_T > \Delta$) while the others are in the long-junction limit ($E_T < \Delta$). Interestingly, for the shortest,  $L$ = 100 nm junctions, a relatively large 100 nA-wide dip of $dV/dI$ is observed in certain fields, even around 8000 G for both Gr/WS$_2$ and Gr/hBN junctions, and oscillates as a function of $B$. In the previous study on graphene ballistic Josephson junctions \cite{BenShalom}, field-dependent and sample-specific differential resistance dips at a low current were also observed around 5000 G, and the $B$ and $V_g$ regions of low $dV/dI$ were termed "superconducting pockets".
In our shortest samples, the superconducting pockets are still visible around $B$ = 16000 G for the Gr/WS$_2$ junction. We note that all Gr/WS$_2$ junctions, even the shortest one with $L$ = 100 nm, are in the diffusive limit because of the shorter $l_e$. 


Whereas the field dependence is similar for both types of 100-nm-long junctions, we find a stark difference for the longer junctions, $L$ = 300 and 500 nm: While superconducting pockets persist around $B$ = 10000 G for Gr/WS$_2$, they are clearly suppressed for Gr/hBN. 
We note that the typical oscillation field scale of the superconducting pockets is about 1.5 G, identical to the width of the Fraunhofer-like pattern main lobe. 


Beyond these $dV/dI$ maps as a function of $I_{\rm dc}$ and $B$ in limited field regions, a broader picture can be obtained by following $dV/dI$ at zero dc current bias (ZBR) over a wide range of $B$. ZBR oscillates between the normal state resistance when no superconductivity is induced and has a minimal $dV/dI$ in the middle of the superconducting pocket when the superconducting proximity effect is strongest. 
Figure 3 shows the ZBR as a function of $B$ for all junctions. For $L$ = 100 nm, the ZBR oscillates with a large amplitude both for Gr/WS$_2$ and Gr/hBN, even near $B$ = 18000 G. On the contrary, for $L$ = 300 nm, oscillations are strongly suppressed, especially for $B$ $>$ 5000 G for Gr/hBN, while they persist for Gr/WS$_2$ even at higher fields. The difference is even more striking for the $L$ = 500 nm junctions. The oscillation amplitudes are considerably different already at a small field, and large oscillations are visible at $B$ = 18000 G for Gr/WS$_2$, while Gr/hBN exhibits almost no oscillations over the entire $B$ range. In \cite{Supp}, we provide the whole data from the $L$=500 nm Gr/WS$_2$ junction, displaying how the oscillations persist up to 70000 G. These results demonstrate that superconducting pockets can persist at much higher fields for Gr/WS$_2$ than for Gr/hBN, in the longest junctions.

We now discuss possible mechanisms by which SOIs can enhance the robustness of the induced superconductivity at high fields. Superconducting pockets at high fields have already been discussed for ballistic junctions \cite{BenShalom}, in terms of Andreev bound states mediated by chaotic ballistic billiard paths localized at the edges of graphene. Those paths can be considered a ballistic analog of the quasiclassical phase-coherent paths and produce mesoscopic fluctuations of the supercurrent $\delta I_c = \sqrt{\langle I_c^2 \rangle - \langle I_c \rangle^2}$ \cite{Altshuler}. In the ballistic short junction limit, $\delta I_c$ is estimated as \cite{Beenakker2}
\begin{equation}
\delta I_c \sim \frac{e \Delta_0}{\hbar},
\end{equation}
where $\Delta_0$ denotes the superconducting gap at $T$ = 0. 
In the diffusive long-junction limit \cite{Altshuler, Houzet},
\begin{equation}
\delta I_c \sim \frac{e E_T}{\hbar}\sqrt{\frac{W}{L}}. 
\end{equation}


We find $\delta I_c \sim$ 240 nA from Eq. (1) expected to be adequate for short Gr/hBN junctions but slightly larger than our experimental value. Reduced experimental values compared to the theoretical ones were already reported \cite{Takayanagi, Doh} and may be attributed to barriers at the normal-metal-superconductor interface, along with electromagnetic noise or finite temperature effects. We then estimate $\delta I_c$ for diffusive junctions by using Eq. (2), and obtain $\delta I_c \sim$ 100 nA and 50 nA for $L$ = 300 and 500 nm, respectively. The latter is in almost perfect agreement with the experimental result, while the former is larger than the experimental value. However, equations Eqs. (1) and (2) were evaluated for zero field. The field dependence of the critical current was recently theoretically investigated in two-dimensional ballistic junctions similar to our samples \cite{Meier}: As $B$ increases, the supercurrent is localized near the edges, and $I_c$ decays faster ($I_c \propto 1/B^2$) than the typical current fluctuations $\delta I_c$ \cite{Meier}. Moreover, those fluctuations can persist up to high fields for edges whose roughness is characterized by a correlation length of the order of or larger than the Fermi wavelength. In these conditions, they find $\delta I_c= \alpha E_{T}/\Phi_0$, independent of field, which corresponds to the current carried by one ballistic channel, with $\alpha = 2 \pi/9 \sqrt{3}$.  
This yields $\delta I_c \sim$ 200 nA for the 100-nm-long Gr/hBN junction, in qualitative agreement with our experimental findings. These high field fluctuations are specific to ballistic junctions and therefore not expected in the diffusive regime.

In order to explain the robust supercurrent signatures that we find in the diffusive Gr/WS$_2$ junctions, it therefore seems necessary to consider the role of SOIs. 
SOIs favor the formation of edge states, epitomized by the topological quantum spin Hall phase. However, edge states can also exist in a nontopological system, and coexist with bulk states of the same energy. Such edge states are in general sensitive to scattering, but some degree of protection against smooth disorder may exist if the edge states are well separated from bulk states in momentum space. Moreover, spin can also provide additional protection if the spins of the edge state and those of the nearby bulk band are opposite. In the case of graphene on WS$_2$, the analysis of weak antilocalization experiments \cite{Wakamura1, Wakamura2, Zihlmann} has shown that the induced SOIs have both a Rashba-type in-plane component and a tenfold larger out-of-plane component, predominantly of the valley Zeeman type probably \cite{Cummings, David}.
The combined effect of these two types of interactions was theoretically shown to generate nontopological edge states along zigzag edges \cite{Frank}. To explore whether such edge states may explain the persistence and oscillations of supercurrent at high fields, we have performed simulations of $I_c$ as a function of $B$ in graphene stripes containing different types of SOIs \cite{Supp}. We find that a supercurrent persists up to higher fields with SOIs than without SOIs, even when disorder is included. 

We now examine the relation between these edge paths and the chiral edge states of the QH regime. The QH regime develops when the mean free path $l_e \gg 2r_c$, where $r_c$ is the cyclotron radius ($r_c = \hbar k_F / eB$) \cite{Amet}. At $V_g$ = 60 V and $B$ = 10000 G, for example, $2 r_c \sim$ 500 nm, thus  chiral edge states do not contribute to the Andreev bound states localized at the edge. We note that this 2$r_c$ value is much larger than $l_e$ = 30 nm of $L$ = 500 nm Gr/WS$_2$ junction. At higher fields, $r_c$ becomes smaller than $l_e$, so that the Landau localization of bulk states and the formation of chiral edge states may become relevant. We observe superconducting pockets even at 70000 G for the Gr/WS$_2$ junction for $L$ = 500 nm, and $\delta I_c \sim$ 30 nA \cite{Supp}. This value is, however, more than 1 order of magnitude larger than $\delta I_c$ reported in the QH regime with a comparable $r_c$ for shorter and narrower junctions with better quality graphene \cite{Amet}. This may indicate that the supercurrent enhancement by the SOIs can also be effective in the QH regime.  

Another effect of SOIs recently suggested theoretically is the generation of spin-triplet supercurrent flowing close to the edge in combination with magnetic field or exchange interaction in superconductor-normal-metal-superconductor junctions \cite{Brataas, Bergeret1, Bergeret2, Bergeret3}. The characteristic confinement length in this case should be the spin-orbit length ($\lambda_{so}$), estimated to be of the order of a few hundred nm for graphene with strong SOIs. Such a large extent would lead to a supercurrent suppression for fields much below the 10000 gauss range, so that this effect cannot explain the strong lateral confinement we observe.
Another interesting possibility, edge supercurrents induced by two-dimensional vortex lattice formation because of Fermi surface warping \cite{Ostroukh}, also seems unlikely, because such Fermi surface warping was not observed in the previous angular-resolved photoemission spectroscopy of similar graphene/WS$_2$ samples \cite{Henck}. 

In conclusion, we have demonstrated robust proximity-induced superconductivity that persists in high magnetic fields for Gr/WS$_2$ Josephson junctions. Compared to the Gr/hBN control samples, all Gr/WS$_2$ junctions (with $L$ between 100 and 500 nm) have lower mobility and are in the diffusive regime. Nevertheless, and most strikingly for longer junctions, superconducting pockets are still observable at 70000 G, whereas they are suppressed for Gr/hBN junctions with the same $L$. We argue that these robust superconducting signatures stem from quasiballistic states confined along the samples edges, stabilized by the SOIs. Because these edge states carry supercurrent at the micrometer scale, one could envisage further investigations, for instance using more elaborate structures for transport measurements, as well as other techniques such as scanning tunneling microscope (STM) or orbital magnetism measurements. Our findings provide important information for progress toward topological superconductvity in which the combined effects of superconductivity and SOIs play crucial roles. 

We gratefully acknowledge very useful discussions by D. Maslov and A. Meszaros, and help with the sample fabrication from A. Assouline, R. Delagrange, F. Parmentier and R. Ribeiro-Palau. This project is financially supported in part by the Marie Sklodowska Curie Individual Fellowships (H2020-MSCAIF-2014-659420); the ANR Grants DIRACFORMAG (ANR-14-CE32-003), MAGMA(ANR-16-CE29-0027-02), and JETS (ANR-16-CE30-0029-01), the Overseas Research Fellowships by the Japan Society for the Promotion of Science (2017-684) and the CNRS.
K.W. and T.T. acknowledge support from the Elemental Strategy Initiative conducted by the MEXT, Japan ,Grant Number JPMXP0112101001,  JSPS KAKENHI Grant Number JP20H00354 and the CREST(JPMJCR15F3), JST.
C. M. is financially supported by the award of the University Research Fellowships Renewals 2017 (UF160539) and the Research Fellows Enhancement Award 2017 (RGF\textbackslash EA\textbackslash 180090) by the UK Royal Society.



\end{document}